\documentclass[a4paper,fleqn,usenatbib,letters]{mnras}
\usepackage[T1]{fontenc}
\usepackage{ae,aecompl}

\usepackage{graphicx}	
\usepackage{amsmath}	
\usepackage{amssymb}	
\usepackage{latexsym}
\usepackage{url}

\def\mnras{MNRAS}
\def\apj{ApJ}
\def\aap{A\&A}
\def\apjl{ApJL}
\def\apjs{ApJS}

\def\aj{AJ}

\def\ga{\sim}

\title[Accreting Transition Discs with large cavieties]{Accreting Transition Discs with large cavities created by  X-ray photoevaporation in C and O depleted discs}

\author[Ercolano, Weber, Owen]{Barbara Ercolano$^{1,2}$\thanks{E-mail: ercolano@usm.lmu.de (BE)}, Michael L. Weber$^{1}$, James E. Owen$^3$\\
$^{1}$Universit\"ats-Sternwarte M\"unchen, Scheinerstr. 1, 81679 M\"unchen, Germany\\
$^{2}$Excellence Cluster Origin and Structure of the Universe,
Boltzmannstr.2, 85748 Garching bei M\"unchen, Germany\\
$^{3}$Astrophysics Group, Imperial College London, Blackett Laboratory, Prince Consort Road, London SW7 2AZ, UK}

\date{Accepted XXX. Received YYY; in original form ZZZ}

\pubyear{2016}

\begin{document}

\label{firstpage}

\pagerange{\pageref{firstpage}--\pageref{lastpage}}

\maketitle

\begin{abstract}

Circumstellar discs with large dust depleted cavities and vigorous accretion 
onto the central star are often considered signposts for (multiple) giant planet formation. In this letter we show that X-ray photoevaporation operating in discs with modest (factors 3-10) gas-phase depletion of Carbon and Oxygen at large radii ($> 15~AU$) yield the inner radius and accretion rates for most of the observed discs, without the need to invoke giant planet formation. We present one-dimensional viscous evolution models of discs affected by X-ray photoevaporation assuming moderate gas-phase depletion of Carbon and Oxygen, well within the range reported by recent observations. Our models use a simplified prescription for scaling the X-ray photoevaporation rates and profiles at different metallicity, and our quantitative result depends on this scaling. While more rigorous hydrodynamical modelling of mass loss profiles at low metallicities is required to constrain the observational parameter space 
that can be explained by our models, the general conclusion that metal 
sequestering at large radii may be responsible for the observed diversity of transition discs is shown to be robust. Gap opening by giant planet formation
may still be responsible for a number of observed transition discs with large 
cavities and very high accretion rate.

\end{abstract}


\section{Introduction}

The circumstellar discs of material left over from the star formation process provide the dust and gas from which planetary systems are formed. 
The environmental conditions for planet formation are thus determined by these planet-forming discs, which evolve and disperse as they give birth to planets. Interestingly, the timescales of disc dispersal are comparable to those of planet formation, suggesting that the dispersal mechanism dominates disc evolution right at the time at which planets form and/or that the formation of planets triggers the dispersal mechanism. Indeed, the planet formation process also strongly affects the disc, making the combined problem of planet formation and disc evolution a strongly coupled and complex problem.
Discs that are believed to be on the verge of dispersal, so-called ``Transition discs'' (TDs), are thus particularly important tools to study the planet formation process, as they can be used to probe different mechanisms that may be at play at this crucial time of disc evolution. 
TDs, are classically identified as discs showing evidence of an (at least partially) evacuated inner dust cavity \citep{1989AJ.....97.1451S}, the size of this cavity can vary from sub-AU sizes to many tens of AU. Most TDs also show evidence of gas accretion onto the central star, with accretion rates sometimes as vigorous as those observed in younger, less evolved planet-forming discs \citep{2014A&A...568A..18M}. Different physical processes may be at play for the formation of TDs (e.g. photoevaporation, MHD processes, dust evolution, planet-disc interactions), in particular disc dispersal models based on photoevaporation from the central star, naturally produce TDs, but they fail to reproduce their full demographics \citep{2011MNRAS.412...13O, 2017RSOS....470114E}. 

TDs with large cavities and vigorous (e.g. of order 10$^{-8}M_{\odot}$/yr or more) accretion  rates are a challenge for current photoevaporation models \citep{2016PASA...33....5O, 2017RSOS....470114E}, and planet-disc interactions are often invoked for the formation of these objects.  However numerical simulations generally fail to reproduce all observables simultaneously \citep{2012ApJ...755....6Z}.  \citet{2011ApJ...738..131D} 
argue that accreting TDs with large inner holes are signposts of young multiplanet systems, but, based on observational constraints on stability and occurence rate, 
\citet{2016ApJ...825...77D} 
argue against the planetary sculpting hypothesis for TDs
An alternative explanation for the large accretion rates measured in large cavity TDs was recently presented by \citet{2017ApJ...835...59W}, involving a magnetically driven supersonic accretion of low surface density material. This model however is at odds with recent observations from the Atacama Large Millimiter Array (ALMA) of the inner cavities of some of these objects, showing that a large quantity of gas is present inside the dust cavity \citep{2014A&A...562A..26B, 2015A&A...579A.106V, 
2016A&A...585A..58V, 2017ApJ...836..201D}. 

Understanding the origins of TDs and their demographics is of fundamental importance if these objects are to be exploited as tools to study the planet formation process. The nature of accreting, large cavity TDs is one of the open challenges at present. 

In this paper we explore the possibility that the recently reported C and O depletion in the gas phase of the outer regions of planet-forming discs  \citep{2017ApJ...842...98D,2011Sci...334..338H, 2013ApJ...776L..38F, 2016ApJ...828...46A, 2016A&A...592A..83K, 2017A&A...599A.113M} may lead to the formation of large cavity, accreting TDs via an enhancement of the X-ray photoevaporation rates at large disc radii. Indeed C and O are major contributors to the opacity in the X-ray,  
a depletion in the gas phase abundance of these elements results in denser gas being heated to temperatures sufficient for escape, resulting in higher gas densities at the sonic point and ultimately more vigorous winds \citep[][EC10]{2010MNRAS.402.2735E}.

The extent to which the demographical parameter space covered by photoevaporation models  can be extended through this mechanism is  explored here by means of one-dimensional viscous evolution models of discs with varying degree of C and O depletion which are subject to X-ray photoevaporation.



\section{Methods}
\subsection{X-ray photoevaporation at low metallicity}

High energy radiation from the central star can thermally unbind disc material which is then centrifugally launched, creating a photoevaporative wind with a surface mass loss rate which can be approximated by $\dot{\Sigma} \sim \rho_b \cdot M_b \cdot c_s$, where $\rho_b$, $M_b$ and $c_s$ are the density, the launch Mach number and the sound speed at the base of the flow.  EC10 show that in the case of X-ray photoevaporation $\rho_b$, and as a second order effect also $c_s$, increase when the gas metallicity, Z, is reduced, leading to higher wind mass loss rates, $\dot{M}_W$, for low metallicity cases. The increase in $\rho_b$ in the low metallicity cases, which is by far the dominant effect, is due to the reduced extinction allowing high density gas at larger columns to be heated to temperatures sufficiently high for the gas to be photoevaporated. Lower metallicity also leads to higher gas temperature, hence an increase in $c_s$, but this effect only plays a secondary role. More quantitatively EC10 find the following relation of wind mass loss rate with metallicity: $\dot{M}_W(Z) \propto Z^{-0.77}$.

This relation is based on detailed thermal and photoionisation calculations, but the models are run in hydrostatic equilibrium and the mass loss rates are estimated from  $\dot{\Sigma} \sim \rho_b \cdot M_b \cdot c_s$, which \citet{2010MNRAS.401.1415O} have shown to underestimate the absolute value of mass loss rates in the case of X-ray photoevaporation.  Full hydrodynamical simulations, based on a scheme that relates the temperature, $T$, to the local properties of the gas via the ionisation parameter, $\xi$, have only been carried out for the solar metallicity case. The form of the $\xi-T$ relation changes at low metallicity, whereby for the same $\xi$ a higher $T$ is achieved, meaning that the gas density at the sonic point is higher. This produced an increase in mass loss rates at low metallicity similar to EC10's predictions. In this work we then adopt the mass loss rates and wind profiles for the Z=1 case presented by \citet{2010MNRAS.401.1415O, 2011MNRAS.411.1104O, 2012MNRAS.422.1880O}
and scale them for lower values of Z using the $\dot{M}_W(Z)$ proportionality.  In practice, we multiply the mass-loss per unit area by the metallicity scaling $Z^{-0.77}$. The resulting profiles for the Z = 1 and 0.1 cases are shown as the red lines in Figure 1.  As the total mechanical luminosity of the flow cannot exceed the absorbed X-ray flux there is a maximum enhancement to the mass-loss rate that depletion can provide. The results of \citet{2011MNRAS.411.1104O} (their Figure 14) suggest a wind energy efficiency of $\sim$8\% at large radius where the depletion will take place. Therefore, we restrict ourselves to a maximum depletion of Z=0.1, noting further depletion will not increase the mass-loss rates and behave similarly to the Z=0.1 model. The approach above relies on the assumption that the global scaling of the mass loss rates with metallicity reported by EC10 also holds locally. To confirm this, we have re-analysed the model grids of EC10 and found that the slope in the $\dot{M}_W(Z)$ proportionality shows moderate variation between values of $\sim$-0.45 and $\sim$-0.87 at radii between 15 and 40~AU. As will be discussed in Section~4, variations in the profile do not affect the conclusions of this work.

\subsection{Evolution of the gas disc}

We consider the viscous evolution of the surface density of a gaseous disc subject to X-ray photoevaporation, which is described by the following equation: 

\begin{equation}
\frac{\partial \Sigma_{\rm g}}{\partial t} = \frac{1}{r} \frac{\partial}{\partial r} \bigg[ 3r^{1/2} \frac{\partial}{\partial r}\big(\nu \Sigma_{\rm g} r^{1/2}\big)\bigg] - \dot{\Sigma}_{\rm W}(r,t).
\label{eq:evo}
\end{equation}

The first term on the right-hand side describes the viscous evolution of the disc \citep{1974MNRAS.168..603L} and the second the mass loss due to photoevaporation \citep[e.g.,][]{2001MNRAS.328..485C}. $\Sigma_{\rm g}$ is the gas disc surface density, $r$ the radial distance from the star in the disc midplane, $\nu$ the kinematic viscosity of the disc, $M_*$ the stellar mass, and $\dot{\Sigma}_W$ the radial photoevaporation profile. For a 0.7 M$_{\odot}$ star, the integrated gas mass loss rate across the disc is $\dot{M}_W= \int 2 \pi r^2 {\dot \Sigma}_{\rm wind} (r)\ \mathrm{d}r \approx 7 \times 10^{-9} M_{\odot} yr^{-1}$ for an X-ray luminosity of 1.1$\cdot 10^{30} erg/sec$.

To solve Equation~(\ref{eq:evo}), we use the 1D viscous evolution code $\textsc{SPOCK}$ detailed in \cite{2015MNRAS.450.3008E}. We discretize Equation~(\ref{eq:evo}) on a grid of 800 points equispaced in $r^{1/2}$ between $0.0025 - 2500$ AU. We assume a disc temperature structure $T \propto r^{-1/2}$, with $T \approx 2100$ K and 4 K at the inner and outer boundaries, respectively. To integrate the viscous term in Equation~(\ref{eq:evo}) we perform a change of variables to transform this into a diffusion equation \citep{1986MNRAS.221..169P}. 
The photoevaporation term is integrated by removing a fixed amount of mass from each cell at every timestep. A floor surface density of $10^{-8}$ g cm$^{-2}$ is used to prevent numerical problems.
The X-ray model has two epochs delineated by the clearing of the hole in the disc, at which point the inner edge of the outer disc is exposed directly to stellar irradiation. Assuming an X-ray penetration depth of $10^{22}$ cm$^{-3}$, we switch to the second epoch once the hydrogen column density is below this value out to the location of the gap. The simulation ends when the surface density is below 0.1 g cm$^{-2}$ at any given point or after 10~Myr. 
To investigate the effects of depletion in the outer disc regions 
of gas-phase C and O we assume a depletion in Z
\footnote{C and O are the main contributors to the opacity in the soft X-ray spectral region, meaning that a depletion in C and O is roughly equivalent to a depletion in Z for what concerns X-ray penetration depth and hence photoevaporative wind rates.} 
for material at disc radii beyond a critical value $R_Z$. 

\subsection{Population syntheses of evolving discs}
We follow \citet{2011MNRAS.411.1104O} and construct population syntheses for the evolving discs by stochastically sampling the X-ray luminosity function of the central stars using observational data for the Taurus region. We use a cumulative X-Ray luminosity function built from observed luminosities in the range $0.3 - 10 keV$ of pre-main sequence stars with masses $0.5 \leq M \leq 1.0 M_{\odot}$ in the Taurus cluster \citep{2007A&A...468..353G}. Following this method, we construct three populations of TDs, with different levels of C and O depletions, characterized by metallicities of Z = 1 (no depletion), 0.3 and 0.1. Each population contains 1000 discs with the same initial mass $M_0 = 0.07 M_{\odot}$, host-star mass $M_*= 0.7 M_{\odot}$ and aspect ratio $h/r = 0.05$ at $R_1$. The Shakura-Sunyaev parameter $\alpha$ and the scaling radius R$_1$ are adjusted in each population in order to match the observed disc lifetimes. 

We tested that for the Z = 1 case we retrieve the TD demographics of Owen et al. (2017), when using the same disc set up ($\alpha = 2.5 \cdot 10^{-3}$, R$_1$ = 18AU). Our subsequent approach for obtaining the final populations from which the TD demographics are drawn, differs from the work of \citep{2011MNRAS.412...13O} in that, as well as sampling the X-ray luminosity function for the central star, we also randomly sample the initial scaling radius, R$_1$, and the critical radius,  R$_Z$, from a uniform distribution. R$_1$ is allowed to vary between 10 and 100~AU, while R$_Z$ varies between 5 and 50~AU. We then choose $\alpha$ accordingly in order to ensure that the dispersal timescale of the populations match observations. We assume a simple linear dependence of $\alpha$ on R$_1$ and R$_Z$ and find that:
\begin{equation}
\alpha = 2 \cdot 10^{-3} + 7.5 \cdot 10^{-5} \cdot R_1, for Z = 1,
\end{equation}
\vspace{-0.5cm}
\begin{equation}
\alpha = 8 \cdot 10^{-4} + 2 \cdot 10^{-6} \cdot R_1 + 2 \cdot 10^{-5} \cdot R_Z, for Z = 0.3
\end{equation}
\vspace{-0.5cm}
\begin{equation}
\alpha = 2 \cdot 10^{-4} + 2 \cdot 10^{-6} \cdot R_1 + 2 \cdot 10^{-5} \cdot R_Z, for Z = 0.1.
\end{equation}
This approach is chosen in order to maximise the diversity in the resulting TD demographics, and it is justified by the fact that there is no a-priori reason for  $\alpha$, R$_1$ and R$_Z$ to be fixed in all discs. We note however that we have also analysed population syntheses obtained using fixed values of $\alpha$, R$_1$ and R$_Z$ for all three metallicity cases and found that qualitatively our  results do not change. This demonstrates the robustness of our general results against the choice of initial conditions.



\section{Results}

\subsection{Single disc evolution at low metallicity}
Figure 1 compares the evolution of the surface density distribution of a disc irradiated by a central star with L$_X~=~1.1~\cdot~10^{30}$ erg/sec for metallicities of Z = 1 (left panel) and Z = 0.1 (right panel) and critical radius R$_Z = 15~AU$. This model assumes a scaling radius R$_1$~=~18~AU and $\alpha = 2.5\cdot10^{-3}$ for Z = 1 and $\alpha = 4\cdot10^{-4}$ for Z=0.1. Note that the value of $\alpha$ in the Z~=~1 case matches the value assumed by \citet{2011MNRAS.411.1104O}, while in the Z = 0.1 case the value was adjusted to ensure a similar dispersal timescale, which is in both cases observationally constrained. The evident difference between the two cases is the opening radius of the photoevaporative gap, which occurs at $\sim$1~AU and $\sim$50~AU in the Z = 1 and Z = 0.1 case, respectively. The X-ray photoevaporation profile is characterised by a prominent peak at $\sim$1~AU and a broader profile extending to $\sim$80~AU and a secondary, less prominent peak at $\sim$50~AU. In cases where Z is reduced in the outer disc, the secondary peak at $\sim$50~AU dominates and the gap is opened around that location. The mass loss profiles for the relevant metallicity values are superimposed as the thick red lines in Figure 1.  Since the outer regions of the disc are undergoing viscous expansion the gas velocity is radial outwards rather than inwards. Therefore, the location where the mass-loss first exceeds the radial outwards ''decreation'' rate is at large radius rather than at small radius, as would be the case for radial inward gas flow. A similar effect was identified by Morshimi (2012) in the case of a dead-zone. The location of gap opening is then independent of the R$_Z$ assumed, as long as this is smaller than $\sim$50~AU. 



\begin{figure*}
  \includegraphics[width=\textwidth]{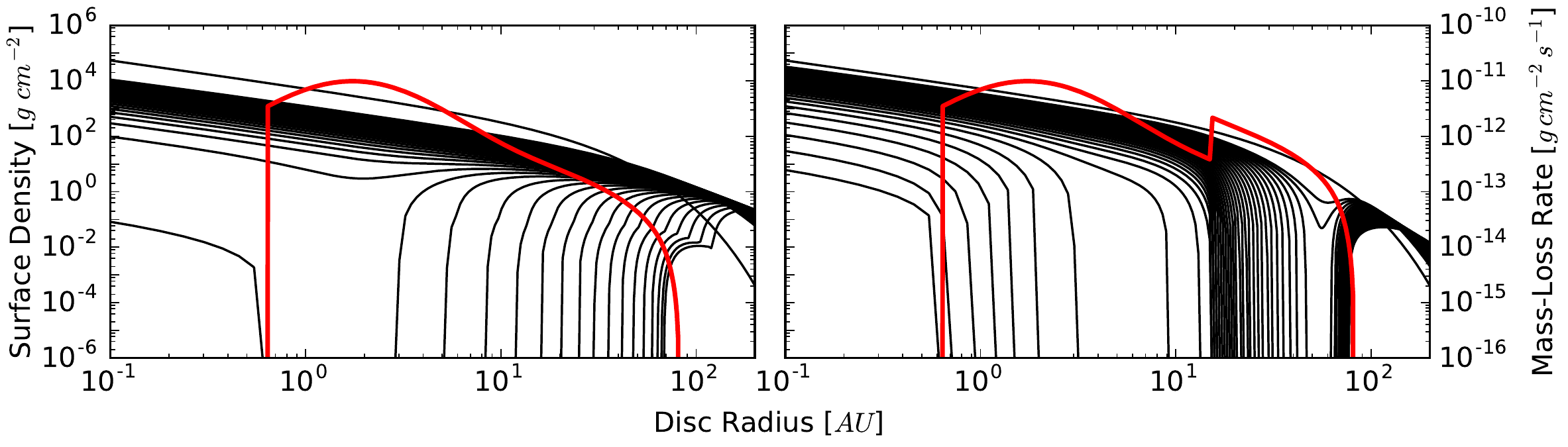}
  \caption{The thin black lines show the surface density evolution for our EUV plus X-Ray driven photoevaporating disc model with the median X-Ray luminosity ($1.1~\cdot~10^{30}erg$ $s^{-1}$) with constant solar metallicity (left) and with a decrease in metallicity by a factor of 10 beyond 15AU (right). The first line shows the surface density at t=0, the following lines show snapshots in 0.1Myr steps, starting at t=1Myr. The thick red lines show the corresponding mass loss profiles. }
  \label{fig:sigmaevol}
\end{figure*}

\subsection{Transition Disc Demographics}
\citet{2011MNRAS.411.1104O} presented population synthesis models of TD demographics and showed that X-ray photoevaporation could not account for all TDs observed at the time. Specifically TDs presenting simultaneously large holes and significant accretion ($\sim10^{-8}M_{\odot}/yr$) lie well outside the parameter space covered by the models, which led the authors to suggest that different mechanisms may be responsible for the cavities observed in these objects \citep{2012MNRAS.426L..96O}.
More recent data has somewhat populated the TD demographics diagram, showing a continuous distribution of sources spanning roughly equally the R$_{in}$-$\dot{M}$ plane. We show all data available to date (excluding G and A-type stars) as grey dots in Figure 2, and refer to the review by \citet{2017RSOS....470114E} for references to the
individual observations. The majority of TDs cannot be explained by the photoevaporation models presented by Owen et al. (2011a) which do not account for C and O depletion in the outer regions of protoplanetary discs. Furthermore Owen et al. (2011, 2017) only assume a single set of initial conditions for all discs (R$_1$~=~18\,AU, $\alpha = 2.5 \cdot 10^{-3}$).

In Figure~2 we show the TD demographics resulting from our population synthesis models which allow a range of initial disc conditions, as described in the previous section. In the three panels we show populations with Z = 1 (panel a), 0.3 (panel b) and 0.1 (panel c) as proxies of similar C and O gas phase depletion in the outer disc. It appears that the number of TDs that can be explained by simple X-ray photoevaporation models is drastically extended already when a factor $\sim$3 depletion in C and O is assumed. There are still a number of strong accretors that are not covered by our model demographics. These objects tend to have large measured disc masses (colour-coded blue by \citet{2017RSOS....470114E}), hinting at cavities having been carved into young discs. 

Another problem of the "standard", Z = 1, X-ray photoevaporation model is the prediction of a large population of discs with large cavities and no accretion signature, which has until now not been matched by observations. This is clearly visible as the red line at the lowest $\dot{M}$ value in Figure 2.a. Indeed the Z = 1 model predicts approximately 98\% of the entire TD population to be non accretors, which is clearly at odds with current observations. The problem is somewhat alleviated when considering models with lower Z. Indeed in the Z = 0.3 and 0.1 models non-accretors account for $\sim$82\% and $\sim$45\% of the whole TD population, respectively. While non-accretors are still not observed in these percentages, this process goes in the right direction to solve the discrepancy with the observations, but other processes also likely contribute to the removal of the outer disc in the remaining cases.

Our simple models, which in the low metallicity case always cause the photoevaporative gap to open at $\sim$50~AU, overpredict the number of TDs that should be observed having cavities of roughly that size. This is particularly evident in the Z = 0.1 case and is an artifact of our simplified approach that scales the photoevaporation rates by a fixed amount for all radii outside a given critical radius, assuming that the normalised mass loss profile remains unchanged. Detailed radiation-hydrodynamical simulations, performed at the relevant metallicities are needed in order to determine the true photoevaporation profile in these cases. Variations in the profile for different values of Z and R$_Z$ would eliminate this feature in the distributions. 

Finally, we note that protoplanetary discs are observed to have different level of C and O depletion. The observed TD demographics are thus likely a result of chemically inhomogeneous populations, i.e. a combination of the cases shown in this work.  

\begin{figure}
  \includegraphics[width=0.48\textwidth]{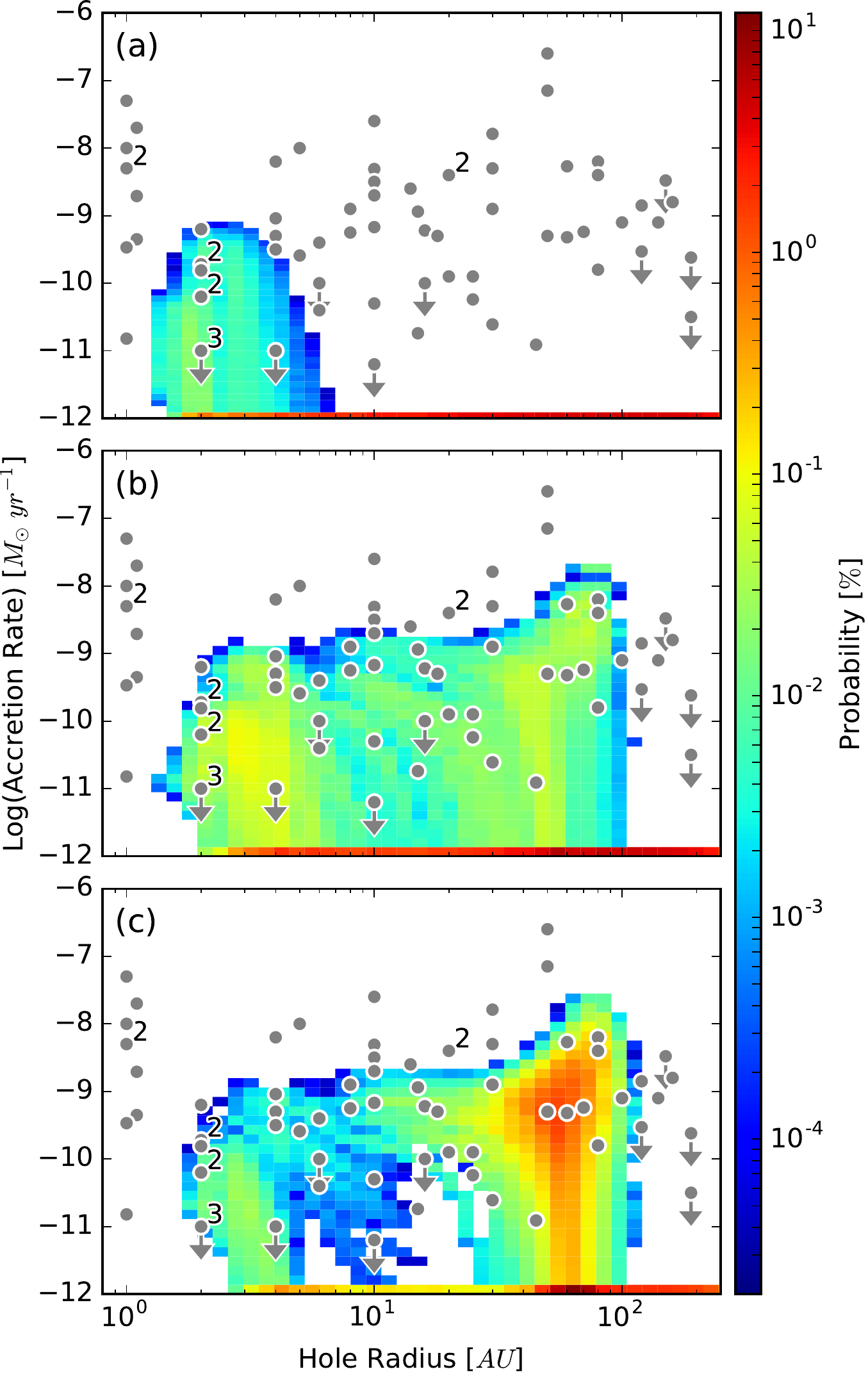}
  \caption{Mass accretion rate versus disc hole size. The grey circles show TDs. Numbers next to the circle indicate the total number of sources at that point, arrows indicate that the accretion rate is an upper limit. The colored areas show the probability to find a TD with the corresponding accretion rate and hole radius, calculated from population synthesis' of our EUV- plus X-Ray driven photoevaporating disc models. Discs with an accretion rate lower than $10^{-12}M_{\odot}yr^{-1}$ are shown at the bottom.  Panel (a) shows our results for disc models with constant solar metallicity, panels (b) and (c) show our results for disc models with a decrease in metallicity by a factor of $\sim$3 (Z = 0.3) and 10 (Z = 0.1).}
  \label{fig:popsynts}
\end{figure}

\section{Conclusions}

We have employed one-dimensional viscuous evolution models of protoplanetary discs affected by X-ray photoevaporation from the central star to investigate the effect of C and O depletion in the outer disc regions on the TD demographics. By considering disc populations with different metallicity, where the X-ray luminosity, the initial disc parameters and the critical radius beyond which the disc is depleted from metals are stochastically sampled we show that we can explain the inner radii and accretion rates for the majority of observed TDs. Populations of C and O depletion in the outer disc also alleviate the discrepancy in the predicted number of non-accreting TDs, which are rarely observed.
On the contrary disc populations with variable initial conditions but no C and O depletion can only explain a small fraction of the observed TDs, namely only those with inner radii of a few AU (less than 10 AU) and mass accretion rates less than 10$^{-9} M_{\odot} yr^{-1}$. Furthermore populations with no C and O depletion predict 98\% of observed TDs to be non-accretors with cavities larger than 10~AU, which is at stark contrast with the observational data. We thus conclude that the recently reported observations showing gas-phase C and O depletion in the outer regions of protoplanetary discs may provide a natural explanation to the observed diversity of TDs when considering X-ray photoevapoaration as the main formation route, without the need to invoke giant planet formation.

A number of important questions however remain that cannot be answered by our simple models. The main technical uncertainty is in the photoevaporation profile at different metallicity, for which our model use the simple scaling provided by \citet{2010MNRAS.402.2735E}. We have tested the robustness of our results to the details of the mass loss profiles, by adopting variable scalings of the local mass loss with metallicity, derived from a re-analysis of the EC10 model grids (see Section~2.1). While we find quantitative differences in the specific probability of observing a TD of a given cavity size and accretion rate, the global distribution of cavity sizes and accretion rates are similar. This supports our general conclusion that metal depletion in the outer regions of protoplanetary discs invariably leads to the formation of accreting TDs with large cavities. Full radiation-hydrodynamic simulations are now crucial to provide reliable mass loss profiles in the case of metallicity gradients, allowing a more quantitative comparison of the models with the observations. Our current results motivate the pursue of such models in a future work.

A further question concerns the feedback of disc dispersal on gas phase C and O depletion. As the disc is eroded from the inside-out by photoevaporation, the outer disc may become warmer, detailed thermochemical models could be useful to predict whether some of the sequestred C and O may be returned to the gas phase during the transition phase. Comparisons of observations C and O depletion in the outer region of primordial and large cavity TDs may also be useful to address this question. 

\section*{Acknowledgements}
We thank the referee R. Alexander for a insightful report. We thank E. van Dishoeck, S. Facchini and N. van der Marel for useful discussion of current observational results. 
This work was funded by the Deutsche Forschungsgemeinschaft (DFG, German Research Foundation) - Ref no. FOR 2634/1. 
This research was supported by the Munich Institute for Astro- and Particle Physics (MIAPP) of the DFG Cluster of Excellence \lq{}Origin and Structure of the Universe.\rq{}
\bibliographystyle{mnras}
\bibliography{references.bib}

\begin{thebibliography}{}
\makeatletter
\relax
\def\mn@urlcharsother{\let\do\@makeother \do\$\do\&\do\#\do\^\do\_\do\%\do\~}
\def\mn@doi{\begingroup\mn@urlcharsother \@ifnextchar [ {\mn@doi@}
  {\mn@doi@[]}}
\def\mn@doi@[#1]#2{\def\@tempa{#1}\ifx\@tempa\@empty \href
  {http://dx.doi.org/#2} {doi:#2}\else \href {http://dx.doi.org/#2} {#1}\fi
  \endgroup}
\def\mn@eprint#1#2{\mn@eprint@#1:#2::\@nil}
\def\mn@eprint@arXiv#1{\href {http://arxiv.org/abs/#1} {{\tt arXiv:#1}}}
\def\mn@eprint@dblp#1{\href {http://dblp.uni-trier.de/rec/bibtex/#1.xml}
  {dblp:#1}}
\def\mn@eprint@#1:#2:#3:#4\@nil{\def\@tempa {#1}\def\@tempb {#2}\def\@tempc
  {#3}\ifx \@tempc \@empty \let \@tempc \@tempb \let \@tempb \@tempa \fi \ifx
  \@tempb \@empty \def\@tempb {arXiv}\fi \@ifundefined
  {mn@eprint@\@tempb}{\@tempb:\@tempc}{\expandafter \expandafter \csname
  mn@eprint@\@tempb\endcsname \expandafter{\@tempc}}}

\bibitem[\protect\citeauthoryear{{Ansdell} et~al.,}{{Ansdell}
  et~al.}{2016}]{2016ApJ...828...46A}
{Ansdell} M.,  et~al., 2016, \mn@doi [\apj] {10.3847/0004-637X/828/1/46}, \href
  {http://adsabs.harvard.edu/abs/2016ApJ...828...46A} {828, 46}

\bibitem[\protect\citeauthoryear{{Bruderer}, {van der Marel}, {van Dishoeck}
  \& {van Kempen}}{{Bruderer} et~al.}{2014}]{2014A&A...562A..26B}
{Bruderer} S.,  {van der Marel} N.,  {van Dishoeck} E.~F.,   {van Kempen}
  T.~A.,  2014, \mn@doi [\aap] {10.1051/0004-6361/201322857}, \href
  {http://adsabs.harvard.edu/abs/2014A%26A...562A..26B} {562, A26}

\bibitem[\protect\citeauthoryear{{Clarke}, {Gendrin}  \& {Sotomayor}}{{Clarke}
  et~al.}{2001}]{2001MNRAS.328..485C}
{Clarke} C.~J.,  {Gendrin} A.,   {Sotomayor} M.,  2001, \mnras, \href
  {http://adsabs.harvard.edu/abs/2001MNRAS.328..485C} {328, 485}

\bibitem[\protect\citeauthoryear{{Dodson-Robinson} \&
  {Salyk}}{{Dodson-Robinson} \& {Salyk}}{2011}]{2011ApJ...738..131D}
{Dodson-Robinson} S.~E.,  {Salyk} C.,  2011, \mn@doi [\apj]
  {10.1088/0004-637X/738/2/131}, \href
  {http://adsabs.harvard.edu/abs/2011ApJ...738..131D} {738, 131}

\bibitem[\protect\citeauthoryear{{Dong} \& {Dawson}}{{Dong} \&
  {Dawson}}{2016}]{2016ApJ...825...77D}
{Dong} R.,  {Dawson} R.,  2016, \mn@doi [\apj] {10.3847/0004-637X/825/1/77},
  \href {http://adsabs.harvard.edu/abs/2016ApJ...825...77D} {825, 77}

\bibitem[\protect\citeauthoryear{{Dong} et~al.,}{{Dong}
  et~al.}{2017}]{2017ApJ...836..201D}
{Dong} R.,  et~al., 2017, \mn@doi [\apj] {10.3847/1538-4357/aa5abf}, \href
  {http://adsabs.harvard.edu/abs/2017ApJ...836..201D} {836, 201}

\bibitem[\protect\citeauthoryear{{Du} et~al.,}{{Du}
  et~al.}{2017}]{2017ApJ...842...98D}
{Du} F.,  et~al., 2017, \mn@doi [\apj] {10.3847/1538-4357/aa70ee}, \href
  {http://adsabs.harvard.edu/abs/2017ApJ...842...98D} {842, 98}

\bibitem[\protect\citeauthoryear{{Ercolano} \& {Clarke}}{{Ercolano} \&
  {Clarke}}{2010}]{2010MNRAS.402.2735E}
{Ercolano} B.,  {Clarke} C.~J.,  2010, \mn@doi [\mnras]
  {10.1111/j.1365-2966.2009.16094.x}, \href
  {http://adsabs.harvard.edu/abs/2010MNRAS.402.2735E} {402, 2735}

\bibitem[\protect\citeauthoryear{{Ercolano} \& {Pascucci}}{{Ercolano} \&
  {Pascucci}}{2017}]{2017RSOS....470114E}
{Ercolano} B.,  {Pascucci} I.,  2017, Royal Society Open Science, \href
  {http://adsabs.harvard.edu/abs/2017RSOS....470114E} {4, 170114}

\bibitem[\protect\citeauthoryear{{Ercolano} \& {Rosotti}}{{Ercolano} \&
  {Rosotti}}{2015}]{2015MNRAS.450.3008E}
{Ercolano} B.,  {Rosotti} G.,  2015, \mn@doi [\mnras] {10.1093/mnras/stv833},
  \href {http://adsabs.harvard.edu/abs/2015MNRAS.450.3008E} {450, 3008}

\bibitem[\protect\citeauthoryear{{Favre}, {Cleeves}, {Bergin}, {Qi}  \&
  {Blake}}{{Favre} et~al.}{2013}]{2013ApJ...776L..38F}
{Favre} C.,  {Cleeves} L.~I.,  {Bergin} E.~A.,  {Qi} C.,   {Blake} G.~A.,
  2013, \mn@doi [\apjl] {10.1088/2041-8205/776/2/L38}, \href
  {http://adsabs.harvard.edu/abs/2013ApJ...776L..38F} {776, L38}

\bibitem[\protect\citeauthoryear{{G{\"u}del} et~al.,}{{G{\"u}del}
  et~al.}{2007}]{2007A&A...468..353G}
{G{\"u}del} M.,  et~al., 2007, \mn@doi [\aap] {10.1051/0004-6361:20065724},
  \href {http://adsabs.harvard.edu/abs/2007A%26A...468..353G} {468, 353}

\bibitem[\protect\citeauthoryear{{Hogerheijde} et~al.,}{{Hogerheijde}
  et~al.}{2011}]{2011Sci...334..338H}
{Hogerheijde} M.~R.,  et~al., 2011, \mn@doi [Science]
  {10.1126/science.1208931}, \href
  {http://adsabs.harvard.edu/abs/2011Sci...334..338H} {334, 338}

\bibitem[\protect\citeauthoryear{{Kama} et~al.,}{{Kama}
  et~al.}{2016}]{2016A&A...592A..83K}
{Kama} M.,  et~al., 2016, \mn@doi [\aap] {10.1051/0004-6361/201526991}, \href
  {http://adsabs.harvard.edu/abs/2016A%26A...592A..83K} {592, A83}

\bibitem[\protect\citeauthoryear{{Lynden-Bell} \& {Pringle}}{{Lynden-Bell} \&
  {Pringle}}{1974}]{1974MNRAS.168..603L}
{Lynden-Bell} D.,  {Pringle} J.~E.,  1974, \mnras, \href
  {http://adsabs.harvard.edu/abs/1974MNRAS.168..603L} {168, 603}

\bibitem[\protect\citeauthoryear{{Manara}, {Testi}, {Natta}, {Rosotti},
  {Benisty}, {Ercolano}  \& {Ricci}}{{Manara}
  et~al.}{2014}]{2014A&A...568A..18M}
{Manara} C.~F.,  {Testi} L.,  {Natta} A.,  {Rosotti} G.,  {Benisty} M.,
  {Ercolano} B.,   {Ricci} L.,  2014, \mn@doi [\aap]
  {10.1051/0004-6361/201323318}, \href
  {http://adsabs.harvard.edu/abs/2014A%26A...568A..18M} {568, A18}

\bibitem[\protect\citeauthoryear{{Miotello} et~al.,}{{Miotello}
  et~al.}{2017}]{2017A&A...599A.113M}
{Miotello} A.,  et~al., 2017, \mn@doi [\aap] {10.1051/0004-6361/201629556},
  \href {http://adsabs.harvard.edu/abs/2017A%26A...599A.113M} {599, A113}

\bibitem[\protect\citeauthoryear{{Owen}}{{Owen}}{2016}]{2016PASA...33....5O}
{Owen} J.~E.,  2016, \mn@doi [\pasa] {10.1017/pasa.2016.2}, \href
  {http://adsabs.harvard.edu/abs/2016PASA...33....5O} {33, e005}

\bibitem[\protect\citeauthoryear{{Owen} \& {Clarke}}{{Owen} \&
  {Clarke}}{2012}]{2012MNRAS.426L..96O}
{Owen} J.~E.,  {Clarke} C.~J.,  2012, \mn@doi [\mnras]
  {10.1111/j.1745-3933.2012.01334.x}, \href
  {http://adsabs.harvard.edu/abs/2012MNRAS.426L..96O} {426, L96}

\bibitem[\protect\citeauthoryear{{Owen}, {Ercolano}, {Clarke}  \&
  {Alexander}}{{Owen} et~al.}{2010}]{2010MNRAS.401.1415O}
{Owen} J.~E.,  {Ercolano} B.,  {Clarke} C.~J.,   {Alexander} R.~D.,  2010,
  \mnras, \href {http://adsabs.harvard.edu/abs/2010MNRAS.401.1415O} {401, 1415}

\bibitem[\protect\citeauthoryear{{Owen}, {Ercolano}  \& {Clarke}}{{Owen}
  et~al.}{2011a}]{2011MNRAS.411.1104O}
{Owen} J.~E.,  {Ercolano} B.,   {Clarke} C.~J.,  2011a, \mnras, \href
  {http://adsabs.harvard.edu/abs/2011MNRAS.411.1104O} {411, 1104}

\bibitem[\protect\citeauthoryear{{Owen}, {Ercolano}  \& {Clarke}}{{Owen}
  et~al.}{2011b}]{2011MNRAS.412...13O}
{Owen} J.~E.,  {Ercolano} B.,   {Clarke} C.~J.,  2011b, \mnras, \href
  {http://adsabs.harvard.edu/abs/2011MNRAS.412...13O} {412, 13}

\bibitem[\protect\citeauthoryear{{Owen}, {Clarke}  \& {Ercolano}}{{Owen}
  et~al.}{2012}]{2012MNRAS.422.1880O}
{Owen} J.~E.,  {Clarke} C.~J.,   {Ercolano} B.,  2012, \mnras, \href
  {http://adsabs.harvard.edu/abs/2012MNRAS.422.1880O} {422, 1880}

\bibitem[\protect\citeauthoryear{{Pringle}, {Verbunt}  \& {Wade}}{{Pringle}
  et~al.}{1986}]{1986MNRAS.221..169P}
{Pringle} J.~E.,  {Verbunt} F.,   {Wade} R.~A.,  1986, \mnras, \href
  {http://adsabs.harvard.edu/abs/1986MNRAS.221..169P} {221, 169}

\bibitem[\protect\citeauthoryear{{Strom}, {Strom}, {Edwards}, {Cabrit}  \&
  {Skrutskie}}{{Strom} et~al.}{1989}]{1989AJ.....97.1451S}
{Strom} K.~M.,  {Strom} S.~E.,  {Edwards} S.,  {Cabrit} S.,   {Skrutskie}
  M.~F.,  1989, \mn@doi [\aj] {10.1086/115085}, \href
  {http://adsabs.harvard.edu/abs/1989AJ.....97.1451S} {97, 1451}

\bibitem[\protect\citeauthoryear{{Wang} \& {Goodman}}{{Wang} \&
  {Goodman}}{2017}]{2017ApJ...835...59W}
{Wang} L.,  {Goodman} J.~J.,  2017, \mn@doi [\apj]
  {10.3847/1538-4357/835/1/59}, \href
  {http://adsabs.harvard.edu/abs/2017ApJ...835...59W} {835, 59}

\bibitem[\protect\citeauthoryear{{Zhu}, {Nelson}, {Dong}, {Espaillat}  \&
  {Hartmann}}{{Zhu} et~al.}{2012}]{2012ApJ...755....6Z}
{Zhu} Z.,  {Nelson} R.~P.,  {Dong} R.,  {Espaillat} C.,   {Hartmann} L.,  2012,
  \mn@doi [\apj] {10.1088/0004-637X/755/1/6}, \href
  {http://adsabs.harvard.edu/abs/2012ApJ...755....6Z} {755, 6}

\bibitem[\protect\citeauthoryear{{van der Marel}, {van Dishoeck}, {Bruderer},
  {P{\'e}rez}  \& {Isella}}{{van der Marel} et~al.}{2015}]{2015A&A...579A.106V}
{van der Marel} N.,  {van Dishoeck} E.~F.,  {Bruderer} S.,  {P{\'e}rez} L.,
  {Isella} A.,  2015, \mn@doi [\aap] {10.1051/0004-6361/201525658}, \href
  {http://adsabs.harvard.edu/abs/2015A%26A...579A.106V} {579, A106}

\bibitem[\protect\citeauthoryear{{van der Marel}, {van Dishoeck}, {Bruderer},
  {Andrews}, {Pontoppidan}, {Herczeg}, {van Kempen}  \& {Miotello}}{{van der
  Marel} et~al.}{2016}]{2016A&A...585A..58V}
{van der Marel} N.,  {van Dishoeck} E.~F.,  {Bruderer} S.,  {Andrews} S.~M.,
  {Pontoppidan} K.~M.,  {Herczeg} G.~J.,  {van Kempen} T.,   {Miotello} A.,
  2016, \mn@doi [\aap] {10.1051/0004-6361/201526988}, \href
  {http://adsabs.harvard.edu/abs/2016A%26A...585A..58V} {585, A58}

\makeatother
\end{thebibliography}

\label{lastpage}

\end{document}